# BODY AS CONTROLLER

Lilac Atassi. Email: lilacatassi@gmail.com

**Abstract**

In the process of developing a new digital music interface, the author faced three questions that have attracted little to no attention in the literature. By tracking body joints, a performer can use body parts to directly control a digital music instrument. An immediate question that follows asks which limb(s) is more effective for the instrument. The next question asks that movement should be measured relative to a particular reference point. And the last question asks about the mathematical form of the mapping function from the movement feature to the sound parameters. This paper attempts to discuss why finding an answer to these questions is worthwhile and to provide possible solutions that require further investigation.

A commonly used sensor technology for body point tracking is a depth camera, which is an alternative to much more costly motion capture systems. For instance, the authors in [8] used a depth camera to track and recognize hand gestures as an interface. The authors in [9] also used a depth camera combined with an inertial measurement unit worn on each hand to track body points. However, the recent advancements in computer vision have enabled body point tracking using a regular camera without the depth sensor. This provides an even more accessible solution compared to depth cameras and motion capture systems.

In the interface that I have designed, the computer vision algorithm is based on PoseNet [7], an artificial neural network that detects 17 body points and returns their coordinates in the image. Using a camera and a laptop, 17 body points are tracked. The position and velocity data at these points is then sent to two separate scripts to generate sound and visuals. From the point locations, the velocity and the relative velocity to the other tracked points can be calculated, as illustrated in Figure 1. This set of positions and velocities are generic enough for most body movement to sound mappings. Translating expressive qualities of body movement into sound expressions is the main motivation behind this project.

As a side note, the main limitation of this measurement technology is that the signal-to-noise ratio is inversely related to the distance of the performer from the camera. A higher resolution camera can alleviate this problem.

Figure 1: The set of features for each detected body point.

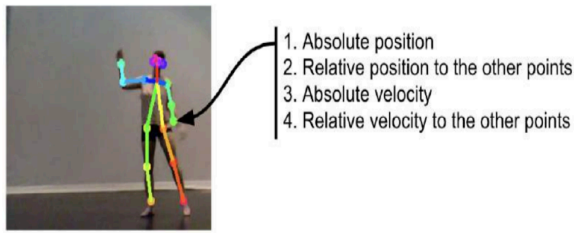

## Body Points to Track

A digital musical instrument requires an input control signal. A diverse set of input signals have been explored by researchers and practitioners. An early example is the hand position for the Theremin. In David Rokeby's *Very Nervous System* [2], the body motion is the input signal. Facial expressions [3], vocal gestures [4], eye movements [5], muscle contraction [6], and brain waves [7] have also been used to control digital musical instruments.

It has been claimed virtuosity is an important factor in the success for the users' appreciation of a new musical instrument. If the performer's body is hidden from the audience, demonstrating virtuosity is impossible [1]. Therefore, a musical interface that uses the performer's body as input control has a great potential for demonstrating virtuosity. This can possibly be explained by embodied simulation, as it has been conjectured as the main enabler of the direct experiential understanding of the intentional and emotional contents of images [11 When we observe a performer performing a musical instrument, our brain simulates the body movement as if we ourselves are performing. The difficulty of the performance therefore is experienced and enjoyed as the result of embodied simulation.

A strong observed correlation between the body motion and the heard sound has also been proposed as an important factor for engaging the audience. Marshall et al. [12] suggest that the perceived liveness of digital music interfaces, which involves the identification of relationships between gestures and resulting sound, impacts on the emotional response of members of the audience.

But these are not guides for which body parts and movements are the most effective in demonstrating virtuosity. For my own interface, I considered homunculus [13] as a guide in designing the musical interface Homunculus, is a figure that looks like a little human, whose form indicates the amount of cortical area dedicated to motor or somatosensory functions of each body part. The body points with a strong correlation between motor homunculus and somatosensory homunculus are perhaps easy for a performer to skillfully control and have heavier weights in embodied simulation, which consequently leads to a stronger virtuosity potential.

In the homunculus, the fingers have the biggest amount of cortical area dedicated to their somatosensory functions. Given the body tracking method I have used for this project cannot detect and track fingers, the wrists and ankles are used to control sound parameters. In an early design of my interface, both wrists and elbows were being tracked and used to control sound parameters. This limited the performer's movements, as the strong correlation between elbow and wrist movement reduced independent control. The question of which body point to track in a musical interface is still an open question that needs more experimentation.

## Reference Point

To track and measure body movement, a reference point is needed. The most commonly used approaches are external physical and virtual reference points. The antennas of the Theremin, for instance, are physical reference points. Virtual reference points are also commonly used, for [10, 16], where the reference point is virtual and is attached to the depth camera used to track body movements.

In my first design, the horizontal center of the camera view was used as a reference point. To let the performer know where the reference point was, the study required a marker on the floor that was aligned with the center point of the camera view. The audience had difficulty observing the relationship between body motion and the virtual reference point.

To mitigate this problem, in the second design, I used the right shoulder point as the reference point for the right wrist location. Using this method, given the arm length of the performer, it is possible to calculate the two end points of the horizontal distance range between the wrist and shoulder. The downside is that, if the distance of the performer to the camera changes, or their orientation with respect to the camera changes, the calculated distance between the wrist and shoulder will be incorrect.

In the third and final design, I used the shoulder-to-shoulder distance as the reference distance, to estimate where the wrist was in the horizontal range. More, the shoulder-to-hip distance was used as the reference distance, to estimate where the wrist was in the vertical range. This means, for instance, that the maximum horizontal right-wrist-to-shoulder distance is set as two times the shoulder-to-shoulder distance to the right side and 1.5 times to the left side. This allows better position estimation when the body is not perpendicular to the camera, without assuming a certain distance between the performer and the camera.

Using the body points as reference points has two possible advantages. First, the performer's body schema (postural model of body) is ready to control the interface with virtually no training time. For instance, without any external feedback, the performer knows how far their right wrist is from their shoulder and from their hip. This allows the performer to move their wrist in that range with a good idea of the distance ratio to the two end points, even when their eyes are closed. Second, using a mapping function that maps the distance ratio between to body points to sound, and showing this during the performance will allow the audience to quickly notice the causal relationship between body point movements relative to the other body points. I am speculating this, as, guessing via embodied simulation, the audience can more easily experience body point movements relative to another body point, versus body movements relative to an external physical reference point, as they are less likely to have a similar experience.

This question calls for further research to obtain a definite answer, and practical comparisons of the options with novice and skilled music performers.

## Natural Mapping

Mapping strategies have been claimed to have a large impact on the possibility of demonstrating virtuosity using a musical instrument [15, 19]. At a lower level, the mathematical function form of the mapping remains an open question.

Is there a mapping function that makes the relation between kinesthetics and sound feel more natural to the performer and audience?

In my first set of experiments, the speed of wrist movement was linearly mapped to sound amplitude, with the slope value designed such that no-movement was mapped to silence, and effortful fast movement was mapped to maximum amplitude. The performers noted the difference in loudness between midrange and maximum movement speed is not easily noticeable. After some experimentation, I came to the conclusion that an exponential function yields a mapping that the performers expected.

The Weber–Fechner law perhaps provides a partial explanation. Non-linear mappings, including logarithmic mapping in Fechner's law, have been traced to the neural structures in the brain [18], providing scientific support for this psychophysical law. For instance, in the Theremin the distance between hand to the pitch antenna of the Theremin is mapped to pitch using an exponential function [17]. In my instrument, the logarithm of the exponential mapping leads to a linear hearing of the loudness which explains the performers' observation in the experiments.

As the input in the present instrument is body movement, the related question is what movements feel natural to the performer when coupled with a certain mapping function and strategy. It is possible the perceived effort of the movement has a large impact on the naturalness of the mapping to both performer and audience. The experiments presented in [14] showed the duration of a movement affects the perceived required effort, and movements towards the midline of the body are less effortful.

Another closely related concept that could benefit the perceived quality of mapping movement to sound is just-noticeable difference (JND). That is, the JND of the input - for instance, the speed of arm movement - should correspond to the JND of the output, such as the sound loudness. If this is not the case, the mapping will feel inconsistent. This requires a more detailed study and experimentation to make a more informed decision when designing the mapping functions.

## Conclusions

In my attempt to save my digital music interface from the "demo and die" syndrome [20], I came across three design decisions that I could not find an answer for in the literature. I presented some answers based on my own experiment, but all three, I believe demand a through investigation to get to more definite answers.